\documentclass[aps,prl,twocolumn,superscriptaddress,10pt]{revtex4-1}
\usepackage[utf8]{inputenc}
\usepackage{amsmath}
\usepackage{amssymb}
\usepackage[normalem]{ulem}
\usepackage{graphicx}
\usepackage{pdfpages}
\makeatletter
\AtBeginDocument{\let\LS@rot\@undefined}
\makeatother

\usepackage{color}
\usepackage{verbatim}
\usepackage{booktabs}
\usepackage{float} 
\usepackage{listings}
\usepackage{alltt}

\usepackage{hyperref}
\usepackage{cleveref}
\usepackage{subfigure}

\usepackage{silence}
\usepackage[normalem]{ulem}

\crefname{equation}{Eq.}{Eqs.}
\Crefname{equation}{Equation}{Equations}
\crefname{figure}{Fig.}{Figs.}
\Crefname{figure}{Figure}{Figures}
\crefname{section}{Sect.}{Sects.}
\Crefname{section}{Section}{Sections}

\newcommand{\ha}{\hat{a}}
\newcommand{\had}{\hat{a}^\dagger}
\newcommand{\hXa}{\hat{X}_A}
\newcommand{\hYa}{\hat{Y}_A}
\newcommand{\hc}{\hat{c}}
\newcommand{\hcd}{\hat{c}^\dagger}

\newcommand{\hbd}{\hat{b}^\dagger}
\newcommand{\hb}{\hat{b}}

\newcommand{\hH}{\hat{H}}
\newcommand{\wc}{\omega_r}

\newcommand{\wB}{\omega_B}

\newcommand{\kA}{\kappa_A}
\newcommand{\kB}{\kappa_B}
\newcommand{\kC}{\kappa_C}

\begin{document}

\title{Itinerant microwave photon detector}
\author{Baptiste Royer}
\affiliation{Institut quantique and D\'epartment de Physique, Universit\'e de Sherbrooke, 2500 boulevard de l'Universit\'e, Sherbrooke, Qu\'ebec J1K 2R1, Canada}
\author{Arne L. Grimsmo}
\affiliation{Institut quantique and D\'epartment de Physique, Universit\'e de Sherbrooke, 2500 boulevard de l'Universit\'e, Sherbrooke, Qu\'ebec J1K 2R1, Canada}
\affiliation{Centre for Engineered Quantum Systems, School of Physics, The University of Sydney, Sydney, Australia}
\author{Alexandre Choquette-Poitevin}
\affiliation{Institut quantique and D\'epartment de Physique, Universit\'e de Sherbrooke, 2500 boulevard de l'Universit\'e, Sherbrooke, Qu\'ebec J1K 2R1, Canada}
\author{Alexandre Blais}
\affiliation{Institut quantique and D\'epartment de Physique, Universit\'e de Sherbrooke, 2500 boulevard de l'Universit\'e, Sherbrooke, Qu\'ebec J1K 2R1, Canada}
%\date{\today}
\affiliation{Canadian Institute for Advanced Research, Toronto, Canada}

%----------------------------------------------------------------------------------------
%	ABSTRACT
%----------------------------------------------------------------------------------------

\begin{abstract}
The realization of a high-efficiency microwave single photon detector is a long-standing problem in the field of microwave quantum optics. Here we propose a quantum non-demolition, high-efficiency photon detector that can readily be implemented in present state-of-the-art circuit quantum electrodynamics. This scheme works in a continuous fashion, gaining information about the arrival time of the photon as well as about its presence.
The key insight that allows to circumvent the usual limitations imposed by measurement back-action is the use of long-lived dark states in a small ensemble of inhomogeneous artificial atoms to increase the interaction time between the photon and the measurement device. 
Using realistic system parameters, we show that large detection fidelities are possible.
\end{abstract}

\maketitle
%----------------------------------------------------------------------------------------
%	ARTICLE
%----------------------------------------------------------------------------------------

\textit{Introduction}\textemdash 
While the detection of localized microwave photons has been realized experimentally \cite{Gleyzes:07a,Johnson:10a,Schuster:07a}, high-efficiency detection of single \emph{itinerant} microwave photons remains an elusive task~\cite{Sathyamoorthy:16a}. Such detectors are increasingly sought-after due to their applications in quantum information processing~\cite{Gisin:07a,Kimble:08a,Narla:16a}, microwave quantum optics~\cite{Gardiner:04a}, quantum radars~\cite{Lloyd:08a,Tan:2008aa,Guha:2009jb}, and even the detection of dark matter axions~\cite{Lamoreaux:13a}.

In recent years, a large number of microwave photon detector proposals have been put forward~\cite{Helmer:09a,Romero:09a,Wong:17a,Kyriienko:16a,Koshino:13a,Koshino:16a,Sathyamoorthy:14a,Fan:14a,Leppakangas:16a,Reiserer:13a}, and some proof-of-principle experiments have been performed~\cite{Chen:11a,Oelsner:17a,Inomata:16a}. 
For their operation, many of these proposals rely on \emph{a priori} information about the arrival time of the photon~\cite{Romero:09a,Wong:17a,Koshino:13a,Reiserer:13a,Inomata:16a}, limiting their applicability.
In this Letter, we will rather be interested in continuous detectors, where the arrival time of a photon can be inferred \emph{a posteriori}~\cite{Kyriienko:16a,Koshino:16a,Sathyamoorthy:14a,Fan:14a,Leppakangas:16a,Helmer:09a,Chen:11a,Oelsner:17a}.
Moreover, we will also focus on non-destructive detection of photons, where photons are \emph{not} destroyed by the measurement device~\cite{Helmer:09a,Sathyamoorthy:14a,Sathyamoorthy:16a,Reiserer:13a}. 
This property proves to be useful in a number of applications, such as quantum networks~\cite{Gisin:07a,Kimble:08a} and the study of quantum measurement~\cite{Wiseman:10a}.
A challenge in designing continuous single photon detectors is set by the quantum Zeno effect, which loosely states that the more strongly a quantum system is measured, the less likely it is to change its state~\cite{Misra:77a,Kraus:81a}. Any non-heralded photon detection scheme based on absorbing the photon into a medium thus faces the problem that strong continuous measurement reduces the absorbtion efficiency, and thus the photon detection efficiency~\cite{Helmer:09a}. 

In this Letter, we introduce a non-destructive and continuous microwave photon detector that circumvents this measurement back-action problem
with minimal device complexity, without requiring any active control pulses, and avoiding the use of non-reciprocal elements~\cite{Sathyamoorthy:14a,Fan:14a}.
In essence, our proposal relies on absorbing a signal photon in a medium made of an ensemble of inhomogeneous artificial atoms, where the presence of long-lived dark states allows to 
increase the effective lifetime of photons inside this composite absorber without lowering its bandwidth
We show that high detection efficiencies can be obtained by weakly and continuously monitoring the ensemble excitation number.
We also present a simple cQED design implementing this idea, where an ensemble of transmon qubits~\cite{Blais:04a,Koch:07a,Wallraff:04a} are continuously measured through standard dispersive measurement.

\begin{figure}[!t]
\centering
\includegraphics[scale = 1.1]{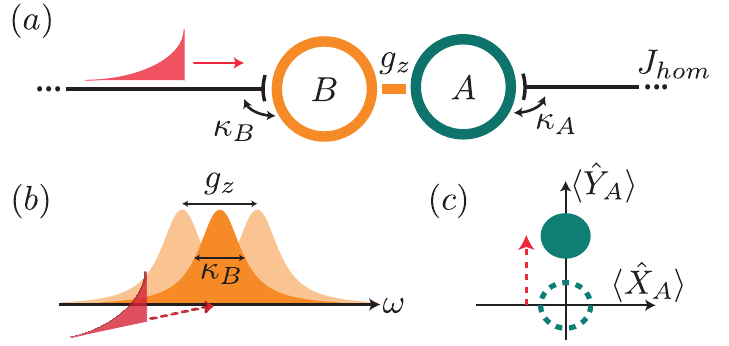}
\caption{(a) Sketch of a single absorber model for photon detection. A signal photon (red) is absorbed in a mode B and induces a coherent state displacement in a harmonic mode A which is measured using standard homodyne measurement. (b) The coupling between A and B induces fluctuations in the lorentzian absorption spectrum of mode B, preventing the absorbtion of incoming photons. (c) Illustration of phase space for mode A  as a photon is absorbed in B.}
\label{fig:schematic}
\end{figure}

\textit{Basic Principle}\textemdash
First consider the toy model illustrated in \cref{fig:schematic}(a), where a signal photon (red) traveling along an input waveguide is absorbed into a single ``absorber'' mode B (orange) at a rate $\kB$. This first mode is coupled to a second ``measurement'' harmonic mode A (green) which decays at a rate $\kA$ into an output port continuously measured using a standard homodyne measurement chain (not shown). In this simple toy model, we assume that the two modes are coupled by the longitudinal interaction ($\hbar = 1$)
\begin{equation}\label{eq:N1Ham}
\hH_I = g_z \hbd\hb (\ha + \had),
\end{equation}
where $\ha$, $\hb$ are the annihilation operators of mode A and B respectively. This interaction implements a textbook photon number measurement: the measured observable $\hbd\hb$ is coupled to the generator of displacement of a pointer state $\hXa = \ha + \had$. As schematically illustrated in \cref{fig:schematic}(c), homodyne measurement of the orthogonal quadrature $\hYa = -i(\ha - \had)$ allows to precisely measure the photon number \emph{inside} the absorber mode B without destroying the photon.

In order to induce a displacement in mode A, a signal photon however needs to first enter the absorber mode B, an unlikely process at large coupling strengths $g_z$. Indeed, as schematically illustrated in \cref{fig:schematic}(b), $\hH_I$ induces quantum fluctuations of the absorber's 
frequency which can prevent it from absorbing the arriving photon. 
In order to minimize this unwanted measurement back-action, the width of these fluctuations, compared with the absorber's linewidth $g_z/\kB$, should ideally be minimized.
On the other hand, the displacement of the measurement mode A, which is given roughly by $g_z/\kB$ as well, should be maximized to improve the detection efficiency~\footnote{The displacement $\sim g_z/\kB$ corresponds to the interaction strength multiplied by the typical lifetime of a photon inside the absorber B.}.  The optimal quantum efficiency of this toy model is obtained by balancing these two conflicting requirements. Numerically we find an optimal operating point at $g_z/\kB = 1$, the smallest coupling strength for which the induced displacement is distinguishable from the vacuum noise $\langle \hYa^2 \rangle_{vacuum} = 1$.

\textit{Numerical Simulations}\textemdash
To model the signal photon arriving at the detector, a source mode C is introduced, with a frequency matching the absorber mode B, $\omega_C = \wB$. To minimize reflection, we take the signal photon linewidth $\kC/\kB = 0.1$ to be much smaller than the absorber's linewidth $\kB$. Following the experiments of Refs.~\cite{Houck:07a,Bozyigit:11a}, this mode is initialized with one excitation leading to a signal photon emission with an exponentially decaying waveform.

The quantum efficiency of this simple photon detector can be determined by numerically simulating multiple realizations of the above scenario and computing the corresponding homodyne current of the measurement mode A. In practice, this is realized by integrating the stochastic master equation~\cite{Wiseman:10a}
\begin{equation}\label{eq:StoME}
\begin{split}
d\rho &= \mathcal L\rho\, dt + \sqrt{\eta_{h}\kA} \mathcal H[-i\ha]\rho\, dW,\\
\hH &= \hH_I - \frac{i\sqrt{\kB \kC}}{2}(\hcd \hb - \hc \hbd),
\end{split}
\end{equation}
where $\hc$ is the annihilation operator of the source mode C and $\mathcal L \bullet$ is the Linbladian superoperator $\mathcal L \bullet = -i[\hH,\bullet] + \sum_j \mathcal D[\hat L_j]\bullet$ with $\hat L_1 = \sqrt{\kA} \ha$, $\hat L_2 = \sqrt{\kB} \hb + \sqrt \kC \hc$. The combination of the term coupling $\hat c$ and $\hat b$ in $\hat H$ and of the composite decay operator $\hat L_2$ assures that the output of mode C is cascaded to the input of mode B~\cite{Gardiner:93a,Carmichael:93a}. Moreover, $\eta_{h}$ is the homodyne measurement chain efficiency, $\mathcal D[\hat L]\bullet = \hat L \bullet \hat L^\dag - \frac{1}{2}\{\hat L^\dag \hat L, \bullet\}$ is the dissipation superoperator and $\mathcal H[\ha]\bullet = \ha \bullet + \bullet \had - \langle \ha + \had \rangle \bullet$ is the homodyne measurement back-action superoperator. The Wiener process $dW$ is a random variable with the statistical properties $E[dW] = 0$ and $E[dW^2] = dt$, where $E[\bullet]$ denotes an ensemble average. 
For each trajectory, the resulting homodyne current is given by
$J_{hom}(t) = \sqrt{\eta_{h}\kA} \langle \hYa \rangle + \xi(t)$,
where $\xi(t) = dW/dt$~\cite{Wiseman:10a}. Here and below, we use ensembles of $N_{traj} = 2000$ trajectories and, to focus solely on the characteristics of the photodetector itself, assume a perfect homodyne detection chain $\eta_{h} = 1$.

For each realization of the homodyne current, we consider that a photon is detected if the convolution of the homodyne signal with a filter, $\bar J_{hom}(t) = J_{hom}(t) \star f(t)$, exceeds a threshold value $Y_{thr}$, i.e.,~if $\text{Max}_t(\bar J_{hom}) > Y_{thr}$. 
To give more weight to times where the signal is, on average, larger, we use $f(t) \propto \langle \hYa(t) \rangle_{ME}$
computed by averaing~\cref{eq:StoME} over all trajectories (equivalently, solving the standard unconditional master equation)~\cite{Fan:14a}.
Given an ensemble of $N_{traj}$ trajectories, the quantum efficiency is then computed as defined in Ref.~\cite{Hadfield:09a}
\begin{equation}
\eta = \frac{N_{click}}{N_{traj}},
\end{equation}
where $N_{click}$ is the number of trajectories where a photon is detected. Although with this model no prior information about the photon arrival time is needed, if this information is available the measurement can be restricted to a time window of length $\tau_m$. In that case, a better metric 
 is the measurement fidelity~\cite{Sathyamoorthy:14a,Koshino:13a}
\begin{equation}
\mathcal F = \frac{1}{2}\left(\eta + 1 - \Gamma_{dark} \times \tau_m\right),
\end{equation}
where $\Gamma_{dark}$ is the dark count rate, i.e. the rate at which the detector ``clicks'' without a signal photon. To maximize the detector repetition rate, $\tau_m$ is set to the smallest value that maximizes the fidelity.% in order to 

For the single absorber model with $g_z/\kB=1$ and $\kA/\kB = 0.2$, we obtain an efficiency of 79\% with $\Gamma_{dark}/\kB = 1.4 \times 10^{-3}$. This translates to a measurement fidelity of $\mathcal F= $ 82\% for a time window of $\kB \tau_m = 125$. The dead time of the detector after a detection event is given by the reset time of the measurement mode A back to vacuum. This corresponds to several decay times $1/\kA$ or, alternatively, can be significantly speed-up by using active cavity reset approaches~\cite{McClure:16a,Bultink:16a,Boutin:16a}.

This scheme is similar to previously studied models~\cite{Helmer:09a,Fan:14a,Milburn:15a} and, although it leads to relatively large detection fidelities, the resulting displacement of mode A is small, $\langle \hYa \rangle \sim g_z/\kB = 1$.
In this situation, adding an imperfect homodyne measurement chain, $\eta_{h} < 1$, will lead to a significant reduction of the quantum efficiency.

\textit{Atom Ensemble}\textemdash
As already pointed out, the key issue with using a single absorber is that both the total displacement of the measurement mode A and the measurement back-action on B scale with $g_z/\kB$. This is a direct consequence of the fact that the time spent in a simple resonant system is given by the inverse of its bandwidth.
In order to increase the quantum efficiency, we thus present a scheme where
the interaction time with the photon is increased while keeping the ratio $g_z/\kB$ constant.

\begin{figure}[!t]
\centering
\includegraphics[scale = 1.1]{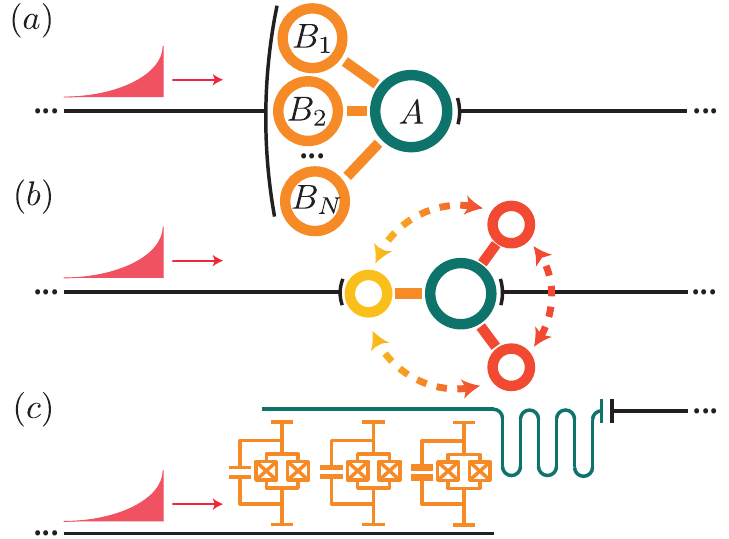}
\caption{(a) The single absorber B is replaced by an ensemble of inhomogeneous modes coupled at the same point of the input waveguide. (b) Redrawing of (a) in the bright and dark states basis for $N=3$. The incoming photon is absorbed into a bright state (yellow) and then passively transferred to dark states (dark orange). (c) Possible circuit QED implementation for $N=3$. Tunable transmon qubits acting as absorbers are coupled capacitively on one side to an input transmission line and on the other side to a measurement resonator.}
\label{fig:schematicN}
\end{figure}

As schematically illustrated in \cref{fig:schematicN}(a), we first replace the single absorber by a small ensemble of $N \lesssim 5$ artificial atoms and, second, we inhomogeneously detune each atom with respect to the average ensemble frequency.
By connecting these absorbers approximately to the same point of the input waveguide~\footnote{In practice, it is sufficient to have the distance $d$ between the artificial atoms to be much smaller than the wavelength of the atoms $d \ll 2\pi v_0/\wB$, where $v_0$ is the speed of light in the waveguide and $\wB$ the transition frequency of the artificial atoms.}, we induce the creation of a superradiant bright state $\hb_+ = 1/\sqrt{N}\sum_i \hb_i$ and dissipationless dark states~\cite{Lalumiere:13a,Loo:13a}. 
Moreover, we assume that the absorbers are coupled to the measurement mode A such that the measured observable is $\hat{N}_B = \sum_i \hbd_i \hb_i$, the total photon number in the ensemble. In this case, the ideal interaction picture Hamiltonian becomes
\begin{equation}\label{eq:Nham}
\hH_I^E = g_z \hat{N}_B \hXa  + \sum_{i=1}^N \Delta_i \hbd_i\hb_i,
\end{equation}
where $\Delta_i = \omega_{Bi} - \wB \lesssim \kappa_{B}$ is the detuning of the $i^{th}$ atom with respect to the average frequency of the ensemble $\wB = \sum_i \omega_{Bi}/N$ and the first term represents the direct generalization of \cref{eq:N1Ham} for an ensemble of atoms.

In this model, an incoming signal photon is absorbed in the collective bright state $\hb_+$ at a rate scaling linearly with $N$. Without loss of generality and to fix the effective collective absorption rate of the detector at $\kB$, we choose the bare linewidth of the atoms to be $\kappa_{Bi} = \kB/N$. In the case where the atoms are on resonance $\Delta_i = 0\, \forall\, i$, the bright and dark subspaces are uncoupled and the model becomes equivalent to the single absorber model illustrated in \cref{fig:schematic}(a)~\cite{Fan:13a}.

On the other hand, non-homogeneous detunings $\Delta_i \neq \Delta_j$ lead to coupling of the bright and dark subspaces. If this coupling is carefully adjusted, a signal photon can then be absorbed into the bright state, transferred to a long-lived dark state and, after some time $\tau_{trap}$, return to the bright state where it is re-emitted. \Cref{fig:schematicN}(b) illustrates this process schematically with the $\hb_+$ bright state (yellow) being coupled to $N-1$ dark states (dark orange). Crucially, changing the detunings affects neither the coupling strength $g_z$ nor the effective linewidth $\kappa_{B}$, which means that the measurement back-action should not be affected either. On the other hand, the total displacement induced in the measurement mode A is changed from $g_z/\kB$ to roughly $g_z \times (1/\kB + \tau_{trap})$. As a result, by increasing $\tau_{trap}$ and reducing $g_z$, we can thus, as desired, significantly increase the quantum efficiency by simultaneously increasing the induced displacement and reducing the measurement back-action. In practice, $\tau_{trap}$ can be made longer by increasing the number of absorbers and optimizing the detunings, $\vec \Delta$, accordingly~\cite{SM}.

\begin{figure}[!t]
\centering
\includegraphics{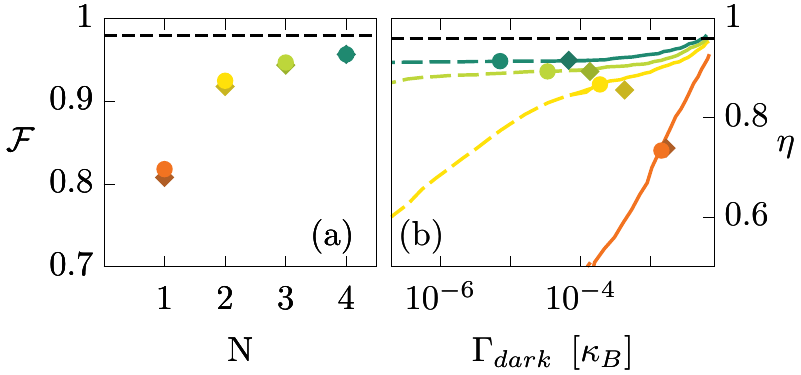}
\caption{(a) Fidelity as a function of the number of absorbers. The full circles are calculated using the ideal model with $\kA/\kB = 0.2$, $g_z^{(1)}/\kB=1$, $g_z^{(2)}/\kB=0.6$, $g_z^{(3)}/\kB=0.5$, $g_z^{(4)}/\kB=0.4$ with the detunings $\vec \Delta^{(2)}/\kB = (0.55,\,-0.55)$, $\vec \Delta^{(3)}/\kB = (0.7,\,-0.7,\, 0)$ and $\vec \Delta^{(4)}/\kB = (0.7,\,-0.7,\, 0.23,\, -0.23)$. The diamonds were calculated using realistic parameters for an ensemble of transmons dispersively coupled to a resonator with $\kB/2\pi = 10$ MHz, $g_z/\chi = 10$ and $T_1,T_2 = 30\, \mu s$. (b) Efficiency of the detector as a function the dark count rate. The solid lines correspond to statistics extracted from trajectories while for the dashed lines $\Gamma_{dark}$ was estimated using an analytical formula. The lines were calculated for the ideal model and the points indicate where the fidelity is maximized. The black dashed line in both panels correspond the upper bound $\eta_{max}$ imposed by the photon shape used here.}
\label{fig:Efficiency}
\end{figure}

We perform full stochastic master equation simulations using \cref{eq:StoME} with the replacements $\hb \rightarrow \hb_+$, $\hH_I \rightarrow \hH_I^E$ and show the increase in measurement fidelity, $\mathcal F$, as a function of ensemble size in \cref{fig:Efficiency}(a). As shown in panel (b), for $N=4$, a quantum efficiency of $\eta = 92\%$ is obtained at a very low estimated dark count rate of $\Gamma_{dark}/\kB = 7 \times 10^{-6}$. For a time window of $\kB \tau_m = 126$ this translates to the measurement fidelity of $\mathcal F = 96\%$ observed in panel (a). As also illustrated in panel (b), it is possible to vary the threshold $Y_{thr}$ to trade a higher dark count rate for a higher efficiency, or the converse. 
Here, the dark count rate $\Gamma_{dark}$ is computed from trajectories with no signal photon (full lines) and, where it is too small to be precisely calculated from trajectories, estimated from time correlations in the filtered signal from vacuum (colored dashed lines)~\cite{SM}.

Importantly, due to the increased interaction time, the measured homodyne signal increases with $N$ and, for $N=4$, is already much larger than vacuum noise. As a result, the detector becomes increasingly robust to potential imperfections in the homodyne detection chain $\eta_{h} < 1$. We, moreover, expect the quantum efficiency to continue increasing as the number of absorbers is raised above 4. Unfortunately, for $N \geq 5 $,  the required Hilbert space size for numerical simulations is impractically large. We note, however, that at $N=4$ the performance are already close to an expected maximum of $\eta_\text{max} \sim 96\%$ indicated by the black dashed line in panels (a) and (b). This upper bound is due to high frequency components of the signal photon that are directly reflected from the absorber and thus do not lead to a detectable signal in mode A~\cite{SM}. The value of this upper bound is linked to the choice of both detector and signal photon parameters and could be improved upon further optimization.

\begin{figure}[!t]
\centering
\includegraphics[scale = 0.46]{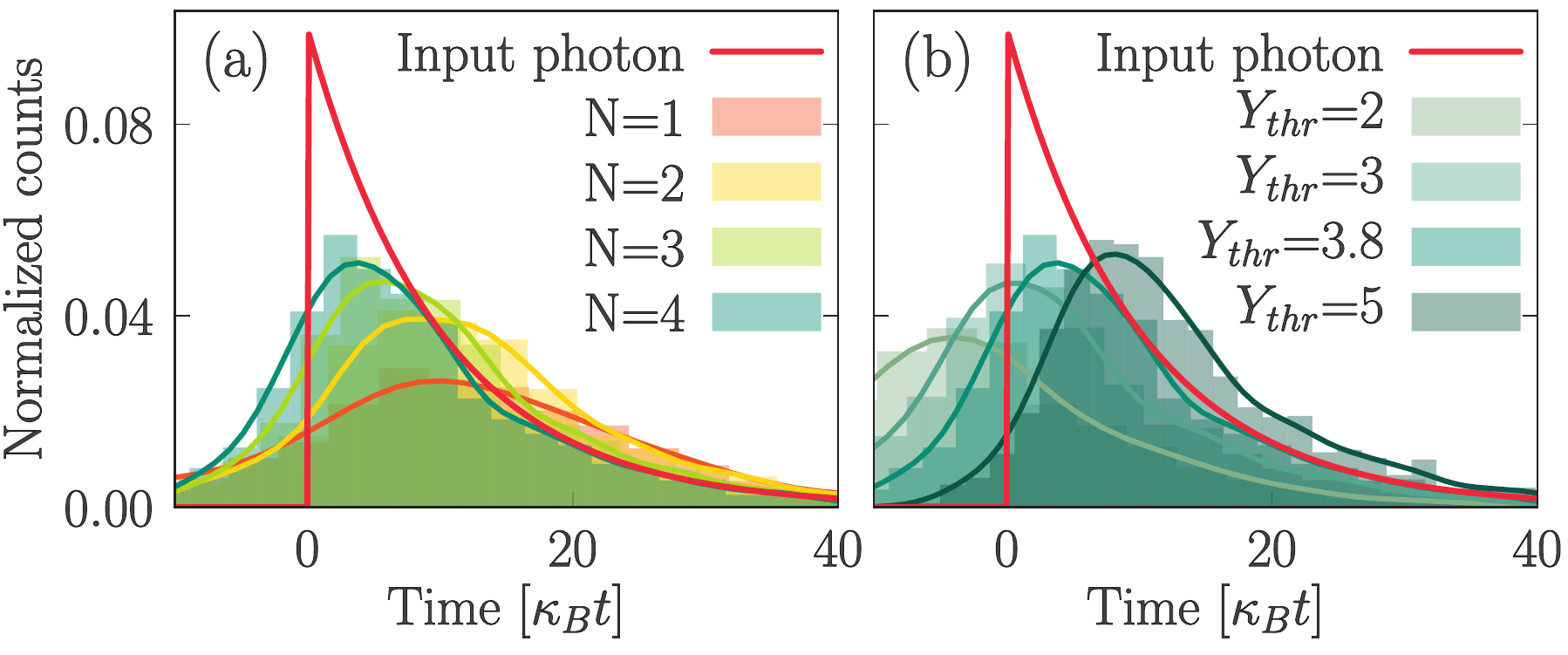}
\caption{(a) Normalized number of detection events as a function of time for different number of absorbers in the ideal model \cref{eq:Nham}. (b) Normalized number of counts for different thresholds for $N=4$. $Y_{thr}=3.8$ is the threshold that maximizes the fidelity in \cref{fig:Efficiency}. In both panels, the input photon shape (red) is shown for comparison and an arbitrary time offset has been substracted from the homodyne signal. }
\label{fig:counts}
\end{figure}

Since our proposal is continuous, the time $\tau_c$ at which the homodyne signal crosses the threshold reveals information about the photon arrival time. \cref{fig:counts} shows histograms of the normalized number of counts for $\tau_c$, as recorded from trajectories where a photon is detected. In \cref{fig:counts}(a), the number of absorbers is varied and the signal threshold, $Y_{thr}$, is set to optimize the fidelity (see \cref{fig:Efficiency}). On the other hand, in \cref{fig:counts}(b), we set $N=4$ and vary the threshold. In both panels, the input photon shape (red) is shown for comparison.
As the threshold increases, the distribution of crossing times narrows and the precision on the arrival time of the photon therefore increases. As mentioned above, increasing $N$ leads to larger homodyne signals. Hence, adding more absorbers allows to increase the threshold which, in turn, improves the arrival time precision.
Moreover, since $1/\kC$ is the longest timescale in these simulations, at $N=4$ the photon shape can be resolved from the histogram.
The mismatch between the distribution and the red line near $\kB t = 0$ is due to the sharp, high frequency feature of the input photon that is reflected from the absorbers without detection.

\textit{Physical implementation}\textemdash
A possible implementation of this model, based on dispersive coupling of transmon qubits, is illustrated in \cref{fig:schematicN}(c). Here, an ensemble of superconducting transmon qubits is capacitively coupled on one side to a transmission line and on the other side to a measurement resonator (mode A). The coupling strength to the resonator is denoted $g$. We take a large detuning between the qubits center frequency $\wB$ and the resonator frequency $\Delta_r = \wc - \wB \gg \kA,\kB,g$ and use the standard dispersive approximation~\cite{SM}.
The absorption of a signal photon by the qubits induces a shift in the resonator frequency which is detected by continuously probing the resonator with a coherent drive corresponding to a field amplitude $\alpha$~\cite{Blais:04a}. In this situation, 
we find that the system of \cref{fig:schematicN}(c) is well described by the displaced dispersive Hamiltonian~\cite{SM}
\begin{equation}\label{eq:longDispersive}
\hH_\chi^D = g_z \hat{N}_B\hXa + \sum_{i=1}^N \Delta_i \hbd_i\hb_i + 2\chi \hat{N}_B \had\ha + \Delta_{+}  \hbd_+ \hb_+,
\end{equation}
where $\chi$ is the usual transmon dispersive shifts~\cite{Koch:07a,SM}, $g_z = 2\chi \alpha$, and $\Delta_+$ results from a combination of the resonator-induced Lamb shift and spurious qubit-qubit coupling~\cite{SM}.
The first two terms correspond exactly to the ideal model Hamiltonian \cref{eq:Nham}, while the two additional last terms are small and imposed by this specific implementation.

For a fixed coupling strength $g_z$, the quantum efficiency is maximized for a small dispersive shift $\chi$ and a large $\alpha$.
However, the dispersive approximation used here is only valid at low photon numbers, imposing an upper bound for the resonator steady state displacement $|\alpha|$. 
As shown by the diamonds in \cref{fig:Efficiency}, working with $\alpha = 5$, we numerically find that the two additional terms in \cref{eq:longDispersive} 
have a minimal impact on the quantum efficiency. Moreover, it is possible to mitigate the detrimental effect of a small $\Delta_+$ by adjusting the detunings $\vec \Delta$.

As an example, choosing realistic parameters $N=4$, $\kB/2\pi = 10 \text{ MHz}$, $\kA/2\pi = 2 \text{ MHz}$, $\chi/2\pi = 0.4 \text{ MHz}$, $\alpha = 5$, $\vec \Delta/2\pi = (6.6,\,-7.4,\,2.3,\,-2.3)\text{ MHz}$ and using state-of-the-art transmon decoherence times $T_1,T_2 = 30\, \mu s$~\cite{McKay:16a}, we obtain $\eta = 92\%$ with $\Gamma_{dark} = 4.2\times 10^{-3}\mu s^{-1}$. Given a time window of $\tau_m = 2\, \mu s$, this corresponds to a large measurement fidelity of $\mathcal F = 96\%$.

\textit{Conclusion}\textemdash
We have presented a high-efficiency, non-destructive scheme for itinerant microwave photon detection where no prior information about the arrival time of the photon is needed. This scheme is based on the continuous measurement of the photon number in an ensemble of inhomogeneous artificial atoms where the photon can be stored for long times due to the existence of long-lived dark states. We also presented a realistic physical implementation of this idea using an ensemble of transmon qubits dispersively coupled to a single resonator. 
Using only four transmons, we estimate that fidelities as high as 96\% are attainable for the photon shape considered here.

Given that the output signal is proportional to the total number of photons inside the absorbers, the same model could potentially to be used as a photon-number resolving detector. Future work will investigate this possibility.
Finally, we note that the same scheme could be applicable to non-destructive detection of single itinerant phonons by coupling the transmons to surface acoustic waves~\cite{Gustafsson:14a,Manenti:17a}.

\begin{acknowledgments}
We thank J\'er\^ome Bourassa, Nicolas Didier for suggesting this project
and St\'ephane Virally for useful discussions.
Part of this work was supported by the Army Research Office under Grant No. W911NF-14-1-0078 and NSERC. This research was undertaken thanks in part to funding from the Canada First Research Excellence Fund and the Vanier Canada Graduate Scholarships.
\end{acknowledgments}

%----------------------------------------------------------------------------------------
%	REFERENCE LIST
%----------------------------------------------------------------------------------------
\bibliographystyle{apsrev4-1}
%\bibliography{biblio}

%

\clearpage


\section*{Supplementary material}

# Supplemental Material for "Itinerant microwave photon detector"

Baptiste Royer,[1] Arne L. Grimsmo,[1, 2] Alexandre Choquette-Poitevin,[1] and Alexandre Blais[1, 3]

[1]*Institut quantique and Départment de Physique, Université de Sherbrooke,*
*2500 boulevard de l'Université, Sherbrooke, Québec J1K 2R1, Canada*
[2]*Centre for Engineered Quantum Systems, School of Physics,*
*The University of Sydney, Sydney, Australia*
[3]*Canadian Institute for Advanced Research, Toronto, Canada*

This supplemental information is organized as follow. In Section I we describe how to compute the shape of an output photon reflected off an ensemble of inhomogeneous modes. The Hamiltonian of the proposed circuit implementation is derived in Sect. II while Sect. III shows that the circuit Hamiltonian is a good approximation of Eq. (6) of the main Letter. Finally, more informations about the numerical simulations are given in Sect. IV.

## I. SHAPE OF REFLECTED PHOTON

In this section, we derive the expression for the shape of a single photon reflected off an ensemble of $N$ modes used to produce Fig. 3 of the main text. In this analysis, we set the measurement back-action from the measurement resonator to zero ($g_z = 0$). This considerably simplifies the calculation and allows us to focus on the photon trapping properties of the system.

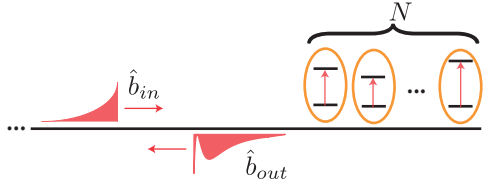

FIG. S1. We consider an ensemble of $N$ modes with different frequencies coupled to a single waveguide, here reprensented as an ensemble of two-level systems without loss of generality. We want to compute the output photon shape $\langle \hat{b}_{out}^\dagger \hat{b}_{out}(t) \rangle$ as a function of the input photon shape $\langle \hat{b}_{in}^\dagger \hat{b}_{in}(t) \rangle$.

Qubit $j$, of transition frequency frequency $\omega_{Bj}$, is described by its lowering $\hat{b}_j$ and raising $\hat{b}_j^\dagger$ operators. Using this notation, the starting point of our analysis are the standard input-output relations, expressed here in a frame rotating at the average qubit frequency $\omega_B = \sum_j \omega_{Bj}/N$ [S1]

$$\hat{b}_{out} = -i \sum_j \sqrt{\kappa_{Bj}} \hat{b}_j + \hat{b}_{in}, \tag{S1}$$

$$\dot{\hat{b}}_j = -i\Delta_j \hat{b}_j - \frac{\kappa_{Bj}}{2} \hat{b}_j - i\sqrt{\kappa_{Bj}} \hat{b}_{in}, \tag{S2}$$

where $\hat{b}_{in}, \hat{b}_{out}$ are respectively the input and ouput fields. For simplicity, we assume that the coupling of each absorber to the input waveguide is identical, $\kappa_{Bj} = \kappa_B/N$. We rewrite Eq. (S2) in matrix form by defining the column vector $\mathbf{b}^T \equiv (\hat{b}_1 \hat{b}_2 \ldots \hat{b}_N)$ and the matrix $\Delta \equiv \text{diag}[\vec{\Delta}]$ with $\vec{\Delta} \equiv (\Delta_1 \Delta_2 \ldots \Delta_N)$,

$$\dot{\mathbf{b}} = -i\Delta \mathbf{b} - \frac{\kappa_B}{2} P_+ \mathbf{b} - i\sqrt{\kappa_B} \hat{b}_{in} \mathbf{e}_+, \tag{S3}$$

where $\mathbf{e}_+^T = 1/\sqrt{N} \times (1\, 1 \ldots 1)$ is the unit vector corresponding to the bright mode $\hat{b}_+ = 1/\sqrt{N} \sum_j \hat{b}_j$ and $P_+ = \mathbf{e}_+ \mathbf{e}_+^\dagger$ is the projector on the subspace spanned by that vector.

It is useful to perform a change of basis, introducing $\tilde{\mathbf{b}} = U\mathbf{b}$ with $U_{jk} = \frac{1}{\sqrt{N}} \exp\left(\frac{jk2\pi}{N}\right)$, such that the dissipative terms take a diagonal form

$$\dot{\tilde{\mathbf{b}}} = -i\tilde{\Delta}\tilde{\mathbf{b}} - \frac{\kappa_B}{2} \widetilde{P}_0 \tilde{\mathbf{b}} - i\sqrt{\kappa_B} \hat{b}_{in} \tilde{\mathbf{e}}_0, \tag{S4}$$

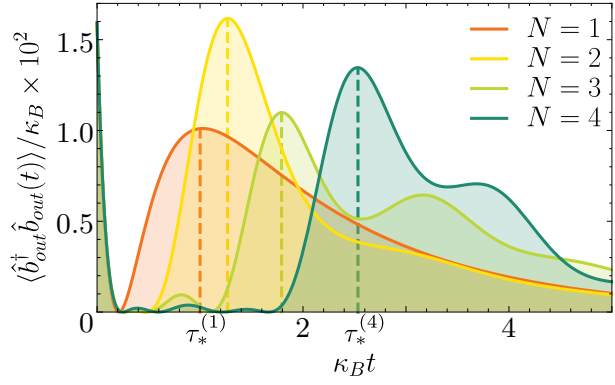

FIG. S2. Output photon shape for different number of absorbers. Here, the input photon shape is a decaying exponential $\langle \hat{b}_{in}^\dagger \hat{b}_{in}(t)\rangle = \kappa_C e^{-\kappa_C t}$ with $\kappa_C/\kappa_B = 0.1$ and, for multiple absorbers, we set detunings to $\vec{\Delta}^{(2)}/\kappa_B = (0.55, -0.55)$, $\vec{\Delta}^{(3)}/\kappa_B = (0.7, -0.7, 0)$ and $\vec{\Delta}^{(4)}/\kappa_B = (0.7, -0.7, 0.23, -0.23)$.

where $\mathrm{U}\mathbf{e}_+ = \tilde{\mathbf{e}}_0$, $\tilde{P}_0 \equiv UP_+ U^\dagger$ and $\tilde{\Delta} \equiv U\Delta U^\dagger$. Projecting on the bright and dark subspaces, this leads to the two coupled Langevin equations

$$\dot{\tilde{b}}_0 = -i\tilde{\mathbf{e}}_0^\dagger \tilde{\Delta} \tilde{P}_D \tilde{\mathbf{b}}_D - i\tilde{\mathbf{e}}_0^\dagger \tilde{\Delta} \tilde{\mathbf{e}}_0 \tilde{b}_0 - \frac{\kappa_B}{2}\tilde{b}_0 - i\sqrt{\kappa_B}\hat{b}_{in},$$
$$\dot{\tilde{\mathbf{b}}}_D = -i\tilde{P}_D \tilde{\Delta} \tilde{P}_D \tilde{\mathbf{b}}_D - i\tilde{P}_D \tilde{\Delta} \tilde{\mathbf{e}}_0 \tilde{b}_0,$$
(S5)

where we have defined $\tilde{b}_0 \equiv \tilde{\mathbf{e}}_0^\dagger \tilde{\mathbf{b}}$ and $\tilde{\mathbf{b}}_D \equiv \tilde{P}_D \tilde{\mathbf{b}}$.

These equations can be solved in Laplace space using the identity $\mathcal{L}[\dot{f}(t)] = s\mathcal{L}[f(t)] - f(0)$, where $\mathcal{L}[f(t)] = \int_0^\infty dt\, e^{-st} f(t)$. Indeed, defining $\tilde{B}_k(s) \equiv \mathcal{L}\left[\tilde{b}_k\right]$, we get

$$\tilde{\mathbf{B}}_D(s) = (s + i\tilde{P}_D \tilde{\Delta} \tilde{P}_D)^{-1}\tilde{\mathbf{b}}_D(0) - i(s + i\tilde{P}_D \tilde{\Delta} \tilde{P}_D)^{-1}\tilde{P}_D \tilde{\Delta} \tilde{\mathbf{e}}_0 \tilde{B}_0(s). \tag{S6}$$

Taking the absorbers to initially be in their ground state, we set $\tilde{\mathbf{b}}_D(0) = 0$ and use the last equation together with Eq. (S5) to solve for $\tilde{B}_0(s)$. To simplify the notation, we define $\delta(s, \Delta) \equiv -i\tilde{\mathbf{e}}_0^\dagger \tilde{\Delta} \tilde{P}_D(s + i\tilde{P}_D \tilde{\Delta} \tilde{P}_D)^{-1} \tilde{P}_D \tilde{\Delta} \tilde{\mathbf{e}}_0$ and find

$$\tilde{B}_0(s) = \frac{-i\sqrt{\kappa_B}}{s + i\tilde{\mathbf{e}}_0^\dagger \tilde{\Delta} \tilde{\mathbf{e}}_0 + i\delta(s,\Delta) + \frac{\kappa_B}{2}} B_{in}(s). \tag{S7}$$

To obtain the shape of the output photon, we replace this result in the Laplace transform of Eq. (S1) and perform the inverse Laplace transform to find

$$\hat{b}_{out}(t) = \mathcal{L}^{-1}\left[\left(1 - \frac{\kappa_B}{s + i\tilde{\mathbf{e}}_0^\dagger \tilde{\Delta} \tilde{\mathbf{e}}_0 + i\delta(s,\Delta) + \frac{\kappa_B}{2}}\right) B_{in}(s)\right]. \tag{S8}$$

Figure S2 was calculated by performing the inverse laplace transform numerically. In our model, the input photon comes from an additional mode with decay rate $\kappa_C$ and initialized in the one photon Fock state $|1\rangle$, such that $B_{in}(s) = \hat{b}_{in}(0)\sqrt{\kappa_C}/(s + \kappa_C/2)$ with $\langle \hat{b}_{in}^\dagger \hat{b}_{in}(0)\rangle = 1$.

To illustrate the increasing trapping time with ensemble size, Fig. S2 shows the shape of the signal photon after re-emission from the detector, $\langle \hat{b}_{out}^\dagger \hat{b}_{out}(t)\rangle$. In the single absorber case $N = 1$ (orange), the photon is absorbed and then re-emitted after a time $\tau_*^{(1)} = 1/\kappa_B$ (dashed orange line). The output photon number also shows, at short times $< 0.2/\kappa_B$, a $\sim 4\%$ component that is directly reflected by the absorber. On the other hand, for $N > 1$ the detunings $\vec{\Delta}$ were optimized such that the photon is re-emitted after an increased time $\tau_*^{(N)} = 1/\kappa_B + \tau_{trap}^{(N)}$ (dashed vertical lines), clearly showing the trapping effect. Also as expected, the photon number at the output is conserved $\int_0^\infty dt\, \langle \hat{b}_{out}^\dagger \hat{b}_{out}(t)\rangle = 1$, corresponding to a non-destructive process. Moreover, because the collective absorption rate has been scaled such that it does not depend on

ensemble size, the small, immediate reflection of about $\sim 4\%$ at short times is identical for all values of $N$ in Fig. S2. Since this component will not lead to a detectable signal in mode A, an upper bound of the quantum efficiency can be obtained from $\eta_{max} \simeq 1 - \int_0^{0.2/\kappa_B} dt \, \langle \hat{b}_{out}^\dagger \hat{b}_{out}(t) \rangle \simeq 96\%$, for these parameters. The value of this upper bound is linked to the choice of both detector and signal photon parameters and could be improved upon further optimization.

## II. CIRCUIT DESIGN

In the following two sections, we derive the Hamiltonian Eq. (6) of the main Letter starting from the circuit illustrated in Fig. S3. In this section, we start from the circuit Lagrangian and perform a Legendre transform to obtain the circuit Hamiltonian. Then, in Sect. III, we show how the circuit Hamiltonian Eq. (S16) approximates the desired Hamiltonian.

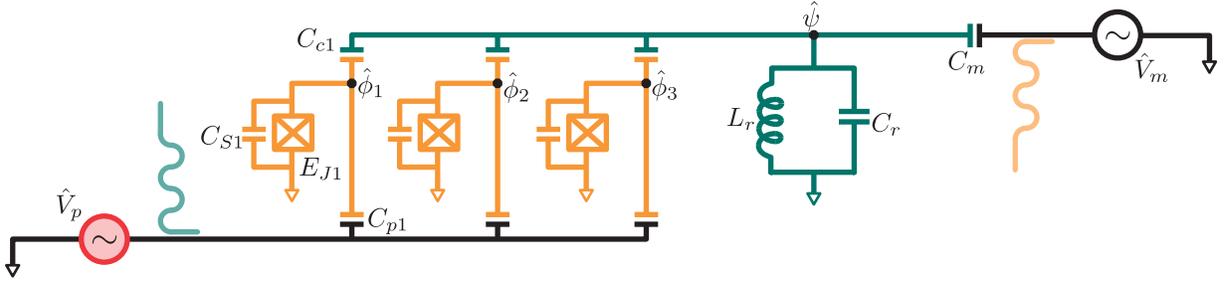

FIG. S3. Circuit design realising the desired Hamiltonian Eq. (6) of the main Letter for $N = 3$. The absorber qubits are in orange ($\hat{\phi}_j$), the measurement resonator is in green ($\hat{\psi}$) and the red voltage $\hat{V}_p$ represents the input photon. The two wavy lines coming out of the transmission lines represent purcell filters that prevent leakage of the modes in the wrong port. The light green purcell filter should thus be at $\omega_r$ and the light orange one at $\omega_B$.

Following standard circuit quantization techniques [S2], the Lagrangian for the circuit illustrated in Fig. S3 takes the form

$$\hat{L} = \frac{C_r \dot{\hat{\psi}}^2}{2} - \frac{\hat{\psi}^2}{2L_r} + \frac{C_m}{2}(\dot{\hat{\psi}} - \hat{V}_m)^2 \\ + \sum_{j=1}^N \frac{C_{cj}}{2}(\dot{\hat{\phi}}_j - \dot{\hat{\psi}})^2 + \frac{C_{Sj}}{2}\dot{\hat{\phi}}_j^2 + \frac{C_{pj}}{2}(\dot{\hat{\phi}}_j - \hat{V}_p)^2 + E_{Jj}\cos\left(\frac{2\pi}{\Phi_0}\hat{\phi}_j\right). \tag{S9}$$

Here, $\hat{\psi}$ represents the readout resonator (mode A) while the $\hat{\phi}_n$ are the qubit phases. It is useful to express the Lagrangian in matrix form with $\vec{\varphi}^T = (\hat{\psi}\,\hat{\phi}_1...\hat{\phi}_N)$, $\hat{L} = \frac{1}{2}\dot{\vec{\varphi}}^T C \dot{\vec{\varphi}} + \vec{v}^T \dot{\vec{\varphi}} - V(\vec{\varphi})$, $V(\vec{\varphi}) = \hat{\psi}^2/2L_r - \sum_j E_{Jj}\cos(2\pi\hat{\phi}_j/\Phi_0)$ and

$$C = \begin{pmatrix} C_r + C_m & -C_{c1} & -C_{c2} & \cdots & -C_{cN} \\ -C_{c1} & C_{S1} + C_{c1} + C_{p1} & 0 & \cdots & 0 \\ -C_{c2} & 0 & C_{S2} + C_{c2} + C_{p2} & \cdots & 0 \\ \vdots & \vdots & \vdots & \ddots & \vdots \\ -C_{cN} & 0 & 0 & \cdots & C_{SN} + C_{cN} + C_{pN} \end{pmatrix}, \tag{S10}$$

$$\vec{v} = \begin{pmatrix} -C_m \hat{V}_m \\ -C_{p1}\hat{V}_p \\ -C_{p2}\hat{V}_p \\ \cdots \\ -C_{pN}\hat{V}_p \end{pmatrix}. \tag{S11}$$

Using this notation, the Hamiltonian is obtained by inverting the capacitance matrix

$$\hat{H} = \frac{1}{2}\vec{q}^T C^{-1} \vec{q} - \vec{v}^T C^{-1} \vec{q} + V(\vec{\varphi}), \tag{S12}$$

where the conjugated variable are given by $\vec{q} = \partial \hat{L}/\partial \dot{\vec{\varphi}} \equiv (\hat{q}_\psi\, \hat{q}_1\, \ldots\, \hat{q}_N)$. Below, we will refer to the element $ij$ of the inverse of the capacitance matrix as $[C^{-1}]_{ij}$, with the index 0 refering to the resonator degree of freedom.

Because a measurement drive is always present on the resonator, it is useful to make the displacement $\hat{V}_m \to V_m(t) + \hat{V}_m$ to separate the classical and the quantum part. For the resonator, we introduce the annihilation and creation operators $\hat{a}, \hat{a}^\dagger$ through

$$\hat{\psi} = \sqrt{\frac{\hbar Z_r}{2}}(\hat{a} + \hat{a}^\dagger), \qquad \hat{q}_\psi = -i\sqrt{\frac{\hbar}{2Z_r}}(\hat{a} - \hat{a}^\dagger), \tag{S13}$$

where $Z_r = \sqrt{L_r [C^{-1}]_{00}}$ is the impedance of the measurement resonator. Moreover, we cast the qubits degree of freedom in the form of a truncated Duffing oscillator with $M$ levels, writing the transition operator of the $j^{th}$ qubit from the $n^{th}$ to the $m^{th}$ level $\hat{\sigma}_{m,n}^{(j)} = |m\rangle\langle n|_j$,

$$\begin{aligned}\hat{\phi}_j &= \sqrt{\frac{1}{2}}\left(\frac{\hbar}{2e}\right)\left(\frac{2E_{Cj}}{E_{Jj}}\right)^{1/4}\sum_{m=0}^{M-1}\sqrt{m+1}(\hat{\sigma}_{m,m+1}^{(j)} + \hat{\sigma}_{m+1,m}^{(j)}),\\ \hat{q}_j &= -ie\sqrt{2}\left(\frac{E_{Jj}}{8E_{Cj}}\right)^{1/4}\sum_{m=0}^{M-1}\sqrt{m+1}(\hat{\sigma}_{m,m+1}^{(j)} - \hat{\sigma}_{m+1,m}^{(j)}),\end{aligned} \tag{S14}$$

where $E_{Cj} = [C^{-1}]_{jj} e^2 / 2$ is the charging energy of the $j^{th}$ qubit. Finally, we introduce the field operators of the measurement transmission line, $\hat{a}_\omega$, and of the input transmission line, $\hat{b}_\omega$, such that

$$\hat{V}_m = \frac{-i}{2}\sqrt{\frac{\hbar Z_m}{\pi}}\int_0^\infty d\omega\,\sqrt{\omega}(\hat{a}_\omega - \hat{a}_\omega^\dagger), \qquad \hat{V}_p = \frac{-i}{2}\sqrt{\frac{\hbar Z_p}{\pi}}\int_0^\infty d\omega\,\sqrt{\omega}(\hat{b}_\omega - \hat{b}_\omega^\dagger). \tag{S15}$$

In these expressions, $Z_m, Z_p$ are respectively the impedance of the measurement and input transmission line. These field operators obey the commutation relations $[\hat{a}_\omega, \hat{a}_{\omega'}^\dagger] = \delta(\omega - \omega')$ and $[\hat{b}_\omega, \hat{b}_{\omega'}^\dagger] = \delta(\omega - \omega')$.

Using Eqs. (S13) to (S15) in Eq. (S12), and performing the standard rotating-wave approximation (RWA) and Born-Markov approximations, the system plus transmission lines Hamiltonian can be expressed as ($\hbar = 1$)

$$\begin{aligned}\hat{H} =\;& \omega_r \hat{a}^\dagger \hat{a} - i\epsilon(t)(\hat{a} - \hat{a}^\dagger) + \sqrt{\frac{\kappa_A}{2\pi}}\int_0^\infty d\omega\,(\hat{a}^\dagger \hat{a}_\omega + \hat{a}\hat{a}_\omega^\dagger) + \int_0^\infty d\omega\,\omega \hat{a}_\omega^\dagger \hat{a}_\omega + \int_0^\infty d\omega\,\omega \hat{b}_\omega^\dagger \hat{b}_\omega\\ &+ \sum_{j=1}^N \sum_{m=0}^{M-1}\left[\omega_{Bj,m}\hat{\sigma}_{m,m}^{(j)} + g_{j,m}(\hat{\sigma}_{m+1,m}^{(j)}\hat{a} + \hat{\sigma}_{m,m+1}^{(j)}\hat{a}^\dagger) + \int_0^\infty d\omega\,\sqrt{\frac{\kappa_{Bj,m}}{2\pi}}(\hat{\sigma}_{m+1,m}^{(j)}\hat{b}_\omega + \hat{\sigma}_{m,m+1}^{(j)}\hat{b}_\omega^\dagger)\right]\\ &+ \sum_{i>j=1}^N \sum_{m=0}^{M-1} J_{ij,m}(\hat{\sigma}_{m+1,m}^{(i)}\hat{\sigma}_{m,m+1}^{(j)} + \hat{\sigma}_{m+1,m}^{(j)}\hat{\sigma}_{m,m+1}^{(j)}).\end{aligned} \tag{S16}$$

In this expressions, the resonator parameters are given by

$$\omega_r = \sqrt{\frac{[C^{-1}]_{00}}{L_r}}, \qquad \epsilon(t) = C_m[C^{-1}]_{00}\frac{1}{\sqrt{2\hbar Z_r}}\times V_m(t), \qquad \kappa_A = C_m[C^{-1}]_{00}\frac{Z_m \omega_r}{4 Z_r}. \tag{S17}$$

We assume that the qubits are in the transmon regime $E_{Jj} \gg E_{Cj}$ and use the results from Ref. [S3] to

obtain the Hamiltonian parameters, which we recall here for completeness

$$\omega_{Bj,m} = \left[-E_{Jj} + \sqrt{8E_{Jj}E_{Cj}}\left(m + \frac{1}{2}\right) - \frac{E_{Cj}}{12}(6m^2 + 6m + 3)\right]/\hbar,$$

$$\kappa_{Bj,m} = C_{pj}\left(\sum_{k=1}^{N}[C^{-1}]_{jk}\right)\frac{\pi Z_p}{R_K}\sqrt{\frac{E_{Jj}}{2E_{Cj}}}(\omega_{Bj,m+1} - \omega_{Bj,m})(m+1),$$

$$g_{j,m} = [C^{-1}]_{0j}\left(\frac{E_{Jj}}{2E_{Cj}}\right)^{1/4}\sqrt{\frac{\pi}{R_K Z_r}}\sqrt{m+1},$$

$$J_{ij,m} = [C^{-1}]_{ij}\left(\frac{E_{Ji}}{2E_{Ci}}\right)^{1/4}\left(\frac{E_{Jj}}{2E_{Cj}}\right)^{1/4}\frac{\pi}{R_K}(m+1),$$

(S18)

where $R_K = h/e^2$ is the resistance quantum. In order to tune *in-situ* the Josephson energy of the junctions, each junction can be replaced by a SQUID, making the Josephson energy dependant on a tunable external flux, $E_{Jj} \to E_{Jj}(\Phi_{ext,j})$.

In the regime where $\omega_r$ is far detuned from $\omega_{Bj,1} - \omega_{Bj,0}$, the two first lines of Eq. (S16) correspond to the desired hamiltonian and, in Sect. III, we show how this Hamiltonian implements the model presented in the Letter. The third line of Eq. (S16) represents a spurious direct coupling between the qubits where the couplings $J_{ij,m}$ are small for the range of parameters used here. Finally, a more complete calculation show that a small spurious coupling between the resonator and the input line voltage is present. This leads to a small, unwanted, decay of the resonator field in the input transmission line. This decay can be negated using a Purcell filter, as illustrated by the light green resonator in Fig. S3. Similarly, a small coupling appears between the qubits and the measurement transmission line, leading to decay of the qubits in the measurement transmission line. This can also be mitigated using a Purcell filter (light orange in Fig. S3).

We note that Eqs. (S1) and (S2) are recovered under a two-level approximation, setting $g_{j,0} = 0$ and identifying $\hat{b}_j = \hat{\sigma}_{0,1}^{(j)}$, $\kappa_{Bj} = \kappa_{Bj,0}$, $\omega_{Bj} = \omega_{Bj,1} - \omega_{Bj,0}$.

### III. DISPERSIVE TRANSFORMATION

In this section, we show how the circuit Hamiltonian Eq. (S16) approximates the desired Hamiltonian, Eq. (6) of the main Letter. First, following standard treatment [S1], the bath degrees of freedom, $\hat{a}_\omega$ and $\hat{b}_\omega$, can be eliminated from Eq. (S16) to obtain the master equation

$$\dot{\rho} = -i[\hat{H}^{ME}, \rho] + \sum_i \mathcal{D}[\hat{L}_i]\rho,$$

(S19)

where

$$\hat{H}^{ME} = \omega_r \hat{a}^\dagger \hat{a} - i\epsilon(t)(\hat{a} - \hat{a}^\dagger) + \sum_{j=1}^{N}\sum_{m=0}^{M-1}\left[\omega_{Bj,m}\hat{\sigma}_{m,m}^{(j)} + g_{j,m}(\hat{\sigma}_{m+1,m}^{(j)}\hat{a} + \hat{\sigma}_{m,m+1}^{(j)}\hat{a}^\dagger)\right]$$

$$+ \sum_{i>j=1}^{N}\sum_{m=0}^{M-1} J_{ij,m}(\hat{\sigma}_{m+1,m}^{(i)}\hat{\sigma}_{m,m+1}^{(j)} + \hat{\sigma}_{m+1,m}^{(j)}\hat{\sigma}_{m,m+1}^{(j)}),$$

$$\hat{L}_1 = \sqrt{\kappa_A}\hat{a},$$

$$\hat{L}_2 = \sum_{j=1}^{N}\sum_{m=0}^{M}\sqrt{\kappa_{Bjm}}\hat{\sigma}_{m,m+1}^{(j)}.$$

(S20)

The first jump operator $\hat{L}_1$ corresponds to the resonator decay into the measurement transmission line while the second jump operator $\hat{L}_2$ to decay of the transmons into the input transmission line. For simplicity, below we assume identical couplings between the transmons and the resonator $g_{j,m} \equiv g_m \,\forall j$ as well as identical transmon-transmon couplings $J_{ij,m} \equiv J_m \,\forall i,j$. Assuming large detuning between the transmons and the measurement resonator, we eliminate their Jaynes-Cummings-like coupling with a Schrieffer-Wolf transformation $\hat{H}^{SW} = \hat{U}^{SW}\hat{H}(\hat{U}^{SW})^\dagger$ with

$$\hat{U}^{SW} = \exp\left[\sum_{m=0}^{M-1}\frac{g_m}{\bar{\omega}_{B(m+1)} - \bar{\omega}_{Bm} - \omega_r}\left(\hat{a}\hat{\sigma}_{m+1,m}^{(+)} - \hat{a}^\dagger\hat{\sigma}_{m,m+1}^{(+)}\right)\right].$$

(S21)

In this expression, we have introduced $\bar{\omega}_{Bm} \equiv \sum_j \omega_{Bj,m}/N$, the average energy of the $m^{th}$ level, and defined $\hat{\sigma}_{m,n}^{(+)} \equiv \sum_j \hat{\sigma}_{m,n}^{(j)}$.

In addition to assuming large transmon-resonator detuning, $|g_m| \ll |\bar{\omega}_{B(m+1)} - \bar{\omega}_{Bm} - \omega_r|$, we set the inhomogeneity of the qubits to be small compared with the resonator-transmon coupling, $|(\omega_{Bj(m+1)} - \omega_{Bjm}) - (\bar{\omega}_{B(m+1)} - \bar{\omega}_{Bm})| \ll |g| \, \forall j$. Moreover, we set the decay rates of all transmons equal and, without loss of generality, we scale their value with the total number $N$ of transmons: $\kappa_{Bj,m} = \kappa_{Bm}/N$. Keeping terms to order $g_m^2/\Delta_{mn}$, where $\Delta_{mn} \equiv \bar{\omega}_{Bm} - \bar{\omega}_{Bn} - \omega_r$, we then find the transformed Hamiltonian

$$\begin{aligned}\hat{H}^{SW} =& \omega_r \hat{a}^\dagger \hat{a} - i\epsilon(t)(\hat{a} - \hat{a}^\dagger) + \sum_{j=1}^{N}\sum_{m=0}^{M} \omega_{Bj,m} \hat{\sigma}_{m,m}^{(j)} + i\epsilon(t) \sum_m \frac{g_m}{\Delta_{m+1,m}}(\hat{\sigma}_{m,m+1}^{(+)} - \hat{\sigma}_{m+1,m}^{(+)}) \\ &+ \sum_m \frac{g_m^2}{\Delta_{m+1,m}} \hat{\sigma}_{m+1,m}^{(+)} \hat{\sigma}_{m,m+1}^{(+)} - \frac{g^2}{\Delta_0} \hat{a}^\dagger \hat{a} \hat{\sigma}_{0,0}^{(+)} + \sum_{mj}\left(\frac{g_m^2}{\Delta_{m+1,m}} - \frac{g_{m+1}^2}{\Delta_{m+2,m+1}}\right)\hat{a}^\dagger \hat{a} \hat{\sigma}_{m+1,m+1}^{(j)} \\ &+ \sum_{m=0}^{M-2} g_m g_{m+1}\left(\frac{1}{\Delta_{m+2,m+1}} - \frac{1}{\Delta_{m+1,m}}\right)\left(\hat{a}\hat{a}\hat{\sigma}_{m+2,m}^{(+)} + \hat{a}^\dagger \hat{a}^\dagger \hat{\sigma}_{m,m+2}^{(+)}\right) \\ &+ \sum_m J_m \hat{\sigma}_{m+1,m}^{(+)} \hat{\sigma}_{m,m+1}^{(+)}\end{aligned} \quad (S22)$$

and the transformed Lindbladians

$$\begin{aligned}\hat{L}_1^{SW} =& \sqrt{\kappa_A}\hat{a} - \sqrt{\kappa_A}\sum_m \frac{g_m}{\Delta_{m+1,m}}\hat{\sigma}_{m,m+1}^{(+)}, \\ \hat{L}_2^{SW} =& \sum_m \sqrt{\frac{\kappa_{Bm}}{N}}\hat{\sigma}_{m,m+1}^{(+)} - \sum_{mj} \sqrt{\frac{\kappa_{Bm}}{N}}\frac{g_m}{\Delta_{m+1,m}}\hat{a}(\hat{\sigma}_{m+1,m+1}^{(+)} - \hat{\sigma}_{m,m}^{(+)}) \\ &- \sum_m \sqrt{\frac{\kappa_{Bm}}{N}}\frac{g_m}{\Delta_{m+1,m}}\hat{a}^\dagger(\hat{\sigma}_{m,m+2}^{(+)} - \hat{\sigma}_{m-1,m+1}^{(+)}).\end{aligned} \quad (S23)$$

A few key observations can significantly reduce the complexity of the above expressions. First, we set the resonator drive at the (pulled) resonator frequency, which means that the induced drive on the transmons is far off-resonant and thus negligible. Second, we consider that a single photon with carrier frequency $\bar{\omega}_{B1} - \bar{\omega}_{B0}$ is sent to the transmons, justifying a two-level approximation for the transmons. In this situation, we can neglect the small, and off-resonant, two-photon transitions. Finally, we note that we neglected a small renormalization of the transmon frequencies $\omega_{Bj,m} - J_m \approx \omega_{Bj,m}$. Defining the normalized modes $\hat{b}_j \equiv \hat{\sigma}_{0,1}^{(j)}$, $\hat{b}_+ \equiv \hat{\sigma}_{0,1}^{(+)}/\sqrt{N}$, the excitation number $\hat{N}_B \equiv \hat{\sigma}_{1,1}^{(+)}$ and using the identity $\hat{\sigma}_{0,0}^{(j)} = 1 - \hat{\sigma}_{1,1}^{(j)}$, the above expressions take the simplified form

$$\hat{H}^{SW} = \tilde{\omega}_r \hat{a}^\dagger \hat{a} - i\epsilon(t)(\hat{a} - \hat{a}^\dagger) + \sum_{j=1}^{N} \omega_{Bj}\hat{b}_j^\dagger \hat{b}_j + N(\chi_{1,0} + J_0)\hat{b}_+^\dagger \hat{b}_+ + 2\left(\chi_{1,0} - \frac{\chi_{2,1}}{2}\right)\hat{N}_B \hat{a}^\dagger \hat{a} \quad (S24)$$

and

$$\begin{aligned}\hat{L}_1^{SW} =& \sqrt{\kappa_A}\hat{a} - \sqrt{\kappa_A}\frac{g\sqrt{N}}{\Delta_{1,0}}\hat{b}_+, \\ \hat{L}_2^{SW} =& \sqrt{\kappa_B}\hat{b}_+ - \sqrt{\kappa_B}\frac{g\sqrt{N}}{\Delta_{1,0}}\hat{a}\left(\frac{2\hat{N}_B}{N} - 1\right),\end{aligned} \quad (S25)$$

where $\chi_{n,m} \equiv g_m^2/\Delta_{n,m}$, $\tilde{\omega}_r \equiv \omega_r - N\chi_{1,0}$ and $\omega_{Bj} \equiv \omega_{Bj1} - \omega_{Bj0}$. We note that here the $\hat{b}$ operators were defined as two-level operators, but since we work in the single excitation subspace (the transmons are excited by a single photon) we can equivalently think of them as ladder operators.

Using these simplified expressions, we now go to a rotating frame for both the resonator and the transmons using the transformation

$$\hat{U}_{rot} = \exp\left\{-it\left[\tilde{\omega}_r \hat{a}^\dagger \hat{a} + [\omega_B + 2\chi(\epsilon/\kappa_A)^2]\hat{N}_B\right]\right\}, \quad (S26)$$

where $\omega_B = \sum_j (\omega_{Bj1} - \omega_{Bj0})/N$ and $\chi \equiv \chi_{1,0} - \chi_{2,1}/2$. Then we take $\epsilon(t) = \epsilon \cos(\tilde{\omega}_r t)$ and neglect fast-rotating terms at $2\tilde{\omega}_r$. Finally, we notice that both jump operators $\hat{L}_1^{SW}, \hat{L}_2^{SW}$ contain terms that rotate a different frequencies. Using the rotating-wave approximation, we neglect the cross terms in the Lindbladian and write

$$\hat{H}_{rot}^{SW} = -i\frac{\epsilon}{2}(\hat{a} - \hat{a}^\dagger) + \sum_{j=1}^{N} \Delta_j \hat{b}_j^\dagger \hat{b}_j + \Delta_+ \hat{b}_+^\dagger \hat{b}_+ + 2\chi \hat{N}_B \hat{a}^\dagger \hat{a} - 2\chi \left(\frac{\epsilon}{\kappa_A}\right)^2 \hat{N}_B, \tag{S27}$$

together with

$$\begin{aligned}
\hat{L}_1^{SW} &= \sqrt{\kappa_A}\,\hat{a}, \\
\hat{L}_2^{SW} &= \sqrt{\kappa_B}\,\hat{b}_+, \\
\hat{L}_3^{SW} &= \sqrt{\kappa_A}\frac{g\sqrt{N}}{\Delta_{1,0}}\hat{b}_+, \\
\hat{L}_4^{SW} &= \sqrt{\kappa_B}\frac{g\sqrt{N}}{\Delta_{1,0}}\hat{a}\left(\frac{2\hat{N}_B}{N} - 1\right),
\end{aligned} \tag{S28}$$

where $\Delta_j \equiv \omega_{Bj1} - \omega_{Bj0} - \omega_B$ and $\Delta_+ = N(\chi_{1,0} + J_0)$. The decay operator $\hat{L}_3^{SW}$ corresponds to a Purcell decay of the bright mode into the measurement transmission line. As already mentioned, this type of decay can be mitigated using standard Purcell filter techniques where the density of states at the transmons frequency is depleted in the measurement transmission line, as illustrated in light orange in Fig. S3. Similarly, $\hat{L}_4^{SW}$ corresponds to a Purcell decay of the measurement resonator into the input transmission line and can also be mitigated by adding another Purcell filter (light green, Fig. S3), depleting the density of states at the resonator frequency in the input transmission line.

Long after the activation of the resonator drive, but before the arrival of a signal photon, the transmons are in their ground state and the resonator is in a coherent steady state $\langle \hat{a} \rangle = \alpha = -\epsilon/\kappa_A$. Following the absorption of a signal photon by the transmons, we are interested in the displacement of the resonator with respect to the average value $\alpha$. Using a displacement transformation $\hat{H}_\chi^D = \hat{D}(\alpha)\hat{H}_{rot}^{SW}\hat{D}^\dagger(\alpha)$, with $\hat{D}(\alpha)\hat{a}\hat{D}^\dagger(\alpha) = \hat{a} - \alpha$, we find the desired Hamiltonian

$$\hat{H}_\chi^D = g_z \hat{N}_B (\hat{a} + \hat{a}^\dagger) + \sum_{j=1}^{N} \Delta_j \hat{b}_j^\dagger \hat{b}_j + 2\chi \hat{N}_B \hat{a}^\dagger \hat{a} + \Delta_+ \hat{b}_+^\dagger \hat{b}_+, \tag{S29}$$

and jump terms

$$\begin{aligned}
\hat{L}_1^D &= \sqrt{\kappa_A}\,\hat{a}, \\
\hat{L}_2^D &= \sqrt{\kappa_B}\,\hat{b}_+,
\end{aligned} \tag{S30}$$

where $g_z = 2\chi\alpha$.

The ideal situation for photodection is to work at large $\alpha$ and small $\chi$. In other words, the ideal situation is reached for a very small dispersive shift probed using a large amplitude coherent state. However, the dispersive transformation is only valid at small photon number $|\alpha|^2 \ll n_{crit}$, limiting the maximal $g_z/\chi$ ratio. Another effect that can in principle significantly reduce the quantum efficiency of the detector is the qubit-induced resonator Kerr non-linearity. At modest $N$ and large detunings, this non-linearity is very small $K = 2N\chi g_0^2/\Delta_{1,0}^2$ and numerical simulations including this effect showed no deviations from the results presented in the main Letter.

## IV. DETAILS ON THE NUMERICAL SIMULATIONS

In this section, we present details concerning the numerical trajectory simulations.

## A. Filtering

As mentionned in the main Letter, the output current is convolved with a filter

$$\bar{J}_{hom}(t) = \int_t^{t+T} d\tau\, J_{hom}(\tau) f(\tau - t), \tag{S31}$$

choosing $T$ so that $f(t > T) \to 0$. As in Ref. [S4], we choose the filter of the same form as the average displacement value computed using a standard master equation simulation. This can be done by, for example, omitting the stochastic part of the stochastic master equation Eq. (2) of the main Letter (by taking $\eta_h = 0$). We denote $y_{av}(t)$ the average displacement calculated this way.

Without loss of generality, we choose to scale the filter so that, at any fixed time, the vacuum noise corresponds to a normal distribution of variance one

$$f(t) = \frac{y_{av}(t)}{\int_0^T d\tau\, y_{av}(\tau)^2}. \tag{S32}$$

This allows to compare the thresholds $Y_{thr}$ for different sets of parameters in a meaningful way.

For high thresholds, the dark count rates are very small and it becomes too numerically expensive to precisely calculate them using trajectories. We therefore derive an approximate analytical formula to compute the dark count rate for high thresholds. First, in the case where there is no signal photon, the homodyne current is given by

$$\bar{J}^0_{hom}(t) = \int_t^\infty d\tau\, f(\tau - t) \xi(\tau), \tag{S33}$$

where $\xi(t)$ is a random variable with statistical properties $E[\xi(t)] = 0$, $E[\xi(t)\xi(t')] = \delta(t - t')$ and the upper bound of the integral has been taken to infinity since $f(t > T) \approx 0$. Due to the above normalization of the filter, the probability that the vacuum signal is above the threshold at any time is given by

$$P(\bar{J}^0_{hom}(t) > Y_{thr}) = \frac{\text{Erfc}(Y_{thr}/\sqrt{2})}{2}. \tag{S34}$$

We define the two-time correlation function for the vacuum homodyne signal

$$C^{(2)}(\tau) = \frac{E[\bar{J}^0_{hom}(0)\bar{J}^0_{hom}(\tau)]}{\int_{-\infty}^\infty d\tau'\, E[\bar{J}^0_{hom}(0)\bar{J}^0_{hom}(\tau')]}, \tag{S35}$$

normalized so that $\int_\infty^\infty d\tau\, C^{(2)}(\tau) = 1$. Using this correlation function, we estimate that the correlation time $\tau_{corr}$ for the vacuum homodyne current is given by

$$\tau_{corr} = \frac{1}{2} \int_{-\infty}^\infty d\tau\, \tau^2 C^{(2)}(\tau). \tag{S36}$$

For a given threshold, the dark count rate can thus be estimated to be

$$\Gamma_{dark} = \frac{P(\bar{J}^0_{hom}(t) > Y_{thr})}{\tau_{corr}}. \tag{S37}$$

In brief, we assume that the dark count rate is given by the probability of a false positive at any time divided by the correlation time of the signal. Figure S4 shows that the estimate Eq. (S37) (dashed lines) approximates well the dark count rate calculated from trajectories (full lines) for moderate dark count rates $5 \times 10^{-5} < \Gamma_{dark} < 10^{-3}$. For smaller dark count rates $\Gamma_{dark} < 5 \times 10^{-5}$, the full lines are unreliable because there are not enough trajectories to calculate precisely the dark count rate. For higher dark count rates $\Gamma_{dark} > 10^{-3}$ the estimate Eq. (S37) is no longer valid: the threshold is so low that the signal can stay above threshold for longer than $\tau_{corr}$. Using Eq. (S37) thus leads to an overestimation of the dark count rate.

## B. Simulations parameters

Tables I and II summarize the parameters used to produce Fig. 3 of the main Letter and Fig. S4. Figure S5 shows the filtered homodyne current for the ideal model Eq. (5) of the main Letter while Fig. S6 shows trajectory results for the more realistic model Eq. (S29).

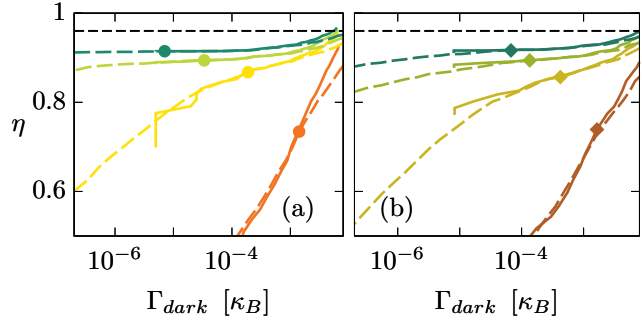

FIG. S4. Efficiency as a function of the dark count rate for the ideal (a) and realistic (b) models. The dark count rate was calculated using trajectories without a signal photon (full lines) and using Eq. (S37) (dashed lines). The points show where the fidelity is maximized.

| $N$ | $\kappa_A$ | $\kappa_C$ | $g_z$ | $\vec{\Delta}$ | $Y_{thr}$ |
|---|---|---|---|---|---|
| 1 | 0.2 | 0.1 | 1 | (0) | 2.2 |
| 2 | 0.2 | 0.1 | 0.6 | (0.55 -0.55) | 2.9 |
| 3 | 0.2 | 0.1 | 0.5 | (0.7 -0.7 0) | 3.4 |
| 4 | 0.2 | 0.1 | 0.4 | (0.7 -0.7 0.23 -0.23) | 3.8 |

TABLE I. Parameters used in the ideal model simulations (Eq. (5) of the main Letter). Here, $\kappa_A, \kappa_C, g_z$ and $\vec{\Delta}$ are in units of $\kappa_B$.

| $N$ | $\kappa_A/2\pi$ [MHz] | $\kappa_{Bi}/2\pi$ [MHz] | $\kappa_C/2\pi$ [MHz] | $g_z/2\pi$ [MHz] | $\chi/2\pi$ [MHz] | $\Delta_+/2\pi$ [MHz] | $\vec{\Delta}/2\pi$ [MHz] | $Y_{thr}$ |
|---|---|---|---|---|---|---|---|---|
| 1 | 2 | 10 | 1 | 10 | 1 | 1 | (0) | 2.1 |
| 2 | 2 | 5 | 1 | 6 | 0.6 | 1.2 | (4.9 -6.1) | 2.7 |
| 3 | 2 | 3.33 | 1 | 5 | 0.5 | 1.5 | (7 -7 0) | 3.0 |
| 4 | 2 | 2.5 | 1 | 4 | 0.4 | 1.6 | (6.6 -7.4 2.3 -2.3) | 3.2 |

TABLE II. Parameters used in the simulations for the more realistic model, Eq. (S29). In all cases, we set intrinsic decoherence times at $T_1 = 30$ $\mu s$ and $T_2 = 30$ $\mu s$.

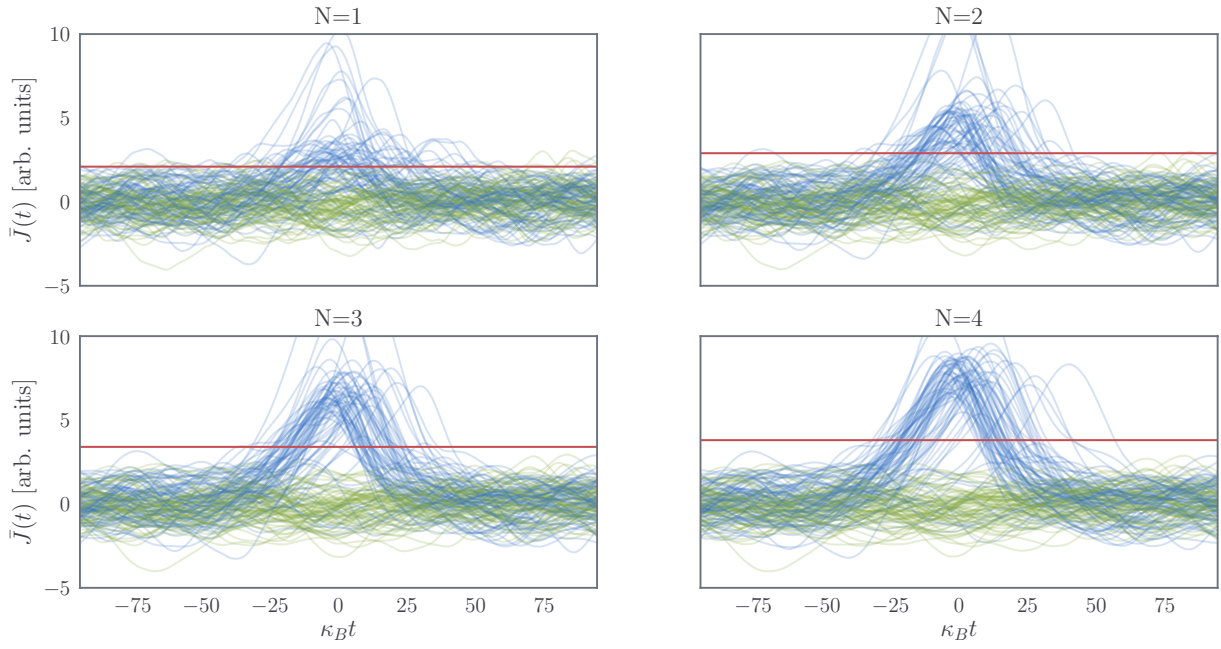

FIG. S5. Filtered homodyne current from 75 trajectories of the ideal model for different number of absorbers with (blue) and without (green) a signal photon. The parameters for each panel are found in Tab. I and the threshold leading to the optimal fidelity is showed in red. The time reference $\kappa_B t = 0$ has been chosen arbitrarily.

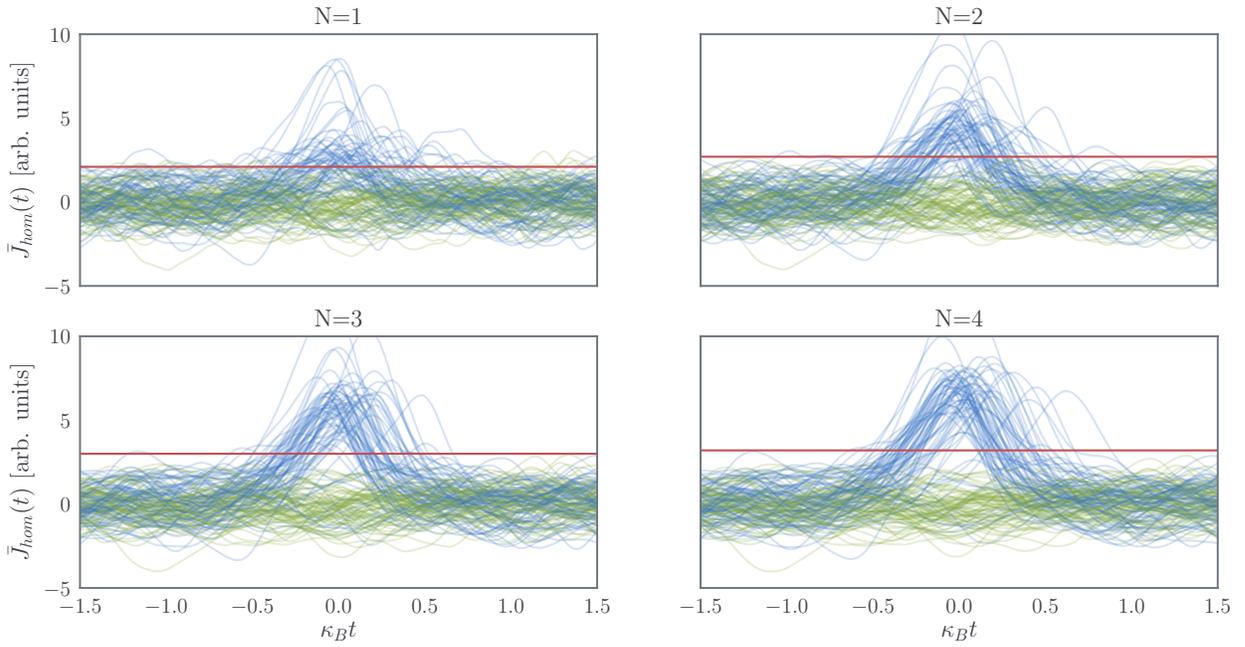

FIG. S6. Filtered homodyne current from 75 trajectories of the realistic model for different number of absorbers with (blue) and without (green) a signal photon. The parameters for each panel are found in Tab. II and the threshold leading to the optimal fidelity is showed in red. The time reference $\kappa_B t = 0$ has been chosen arbitrarily.



\end{document}